\PassOptionsToPackage{unicode}{hyperref}
\PassOptionsToPackage{hyphens}{url}
\PassOptionsToPackage{dvipsnames,svgnames,x11names}{xcolor}
\documentclass[
]{article}
\usepackage{amsmath,amssymb}
\usepackage{lmodern}
\usepackage{iftex}
\ifPDFTeX
  \usepackage[T1]{fontenc}
  \usepackage[utf8]{inputenc}
  \usepackage{textcomp} 
\else 
  \usepackage{unicode-math}
  \defaultfontfeatures{Scale=MatchLowercase}
  \defaultfontfeatures[\rmfamily]{Ligatures=TeX,Scale=1}
\fi
\IfFileExists{upquote.sty}{\usepackage{upquote}}{}
\IfFileExists{microtype.sty}{
  \usepackage[]{microtype}
  \UseMicrotypeSet[protrusion]{basicmath} 
}{}
\makeatletter
\@ifundefined{KOMAClassName}{
  \IfFileExists{parskip.sty}{%
    \usepackage{parskip}
  }{
    \setlength{\parindent}{0pt}
    \setlength{\parskip}{6pt plus 2pt minus 1pt}}
}{
  \KOMAoptions{parskip=half}}
\makeatother
\usepackage{xcolor}
\usepackage{graphicx}
\makeatletter
\def\maxwidth{\ifdim\Gin@nat@width>\linewidth\linewidth\else\Gin@nat@width\fi}
\def\maxheight{\ifdim\Gin@nat@height>\textheight\textheight\else\Gin@nat@height\fi}
\makeatother
\setkeys{Gin}{width=\maxwidth,height=\maxheight,keepaspectratio}
\makeatletter
\def\fps@figure{htbp}
\makeatother
\NewDocumentCommand\citeproctext{}{}
\NewDocumentCommand\citeproc{mm}{%
  \begingroup\def\citeproctext{#2}\cite{#1}\endgroup}
\makeatletter
 \let\@cite@ofmt\@firstofone
 \def\@biblabel#1{}
 \def\@cite#1#2{{#1\if@tempswa , #2\fi}}
\makeatother
\newlength{\cslhangindent}
\newlength{\csllabelwidth}
\newenvironment{CSLReferences}[2] 
 {\begin{list}{}{%
  \setlength{\itemindent}{0pt}
  \setlength{\leftmargin}{0pt}
  \setlength{\parsep}{0pt}
  \ifodd #1
   \setlength{\leftmargin}{\cslhangindent}
   \setlength{\itemindent}{-1\cslhangindent}
  \fi
  \setlength{\itemsep}{#2\baselineskip}}}
 {\end{list}}
\usepackage{calc}

\ifLuaTeX
\usepackage[bidi=basic]{babel}
\else
\usepackage[bidi=default]{babel}
\fi
\babelprovide[main,import]{american}

\def\languageshorthands#1{}
\ifLuaTeX
  \usepackage{selnolig}  
\fi
\IfFileExists{bookmark.sty}{\usepackage{bookmark}}{\usepackage{hyperref}}
\IfFileExists{xurl.sty}{\usepackage{xurl}}{} 
\hypersetup{
  pdftitle={MARTINI: Mock Array Radio Telescope Interferometry of the
Neutral ISM},
  pdfauthor={Kyle A. Oman},
  pdflang={en-US},
  colorlinks=true,
  linkcolor={Maroon},
  filecolor={Maroon},
  citecolor={Blue},
  urlcolor={Blue},
  pdfcreator={LaTeX via pandoc}}

\title{MARTINI: Mock Array Radio Telescope Interferometry of the Neutral
ISM}

\usepackage[affil-it]{authblk}
\usepackage{orcidlink}
\author[1,2%
  \ensuremath\mathparagraph]{Kyle A. Oman%
    \,\orcidlink{0000-0001-9857-7788}\,%
    }

\affil[1]{Institute for Computational Cosmology, Physics Department,
Durham University}
\affil[2]{Centre for Extragalactic Astronomy, Physics Department, Durham
University}
\affil[$\mathparagraph$]{Corresponding author: kyle.a.oman@durham.ac.uk%
}
\date{13 March 2024}

\begin{document}
\maketitle

\section{Summary}\label{summary}

MARTINI is a modular Python package that takes smoothed-particle
hydrodynamics (SPH) simulations of galaxies as input and creates
synthetic spatially- and/or spectrally-resolved observations of the
21-cm radio emission line of atomic hydrogen (data cubes). The various
aspects of the mock-observing process are divided logically into
sub-modules handling the data cube, source galaxy, telescope beam
pattern, noise, spectral model and SPH kernel. MARTINI is
object-oriented: each sub-module provides a class (or classes) which can
be configured as desired. For most sub-modules, base classes are
provided to allow for straightforward customization. Instances of each
sub-module class are given as parameters to an instance of a main
``Martini'' class; a mock observation is then constructed by calling a
handful of functions to execute the desired steps in the mock-observing
process.

\section{Background}\label{background}

The primordial Universe contained predominantly hydrogen, some helium,
and trace amounts of heavier elements. Hydrogen remains the most
abundant element, occuring ubiquitously in stars and
interstellar/intergalactic gas. Atomic hydrogen (i.e.~with an electron
bound to the nucleus) is much more abundant in galaxies where the
gravitational field of dark matter allows gas to collect and cool.
Hydrogen atoms in their lowest energy state exhibit a transition between
the state where the electron spin is aligned and that where it is
anti-aligned with the nuclear spin. The very close (``hyperfine'')
spacing in energy between these two states means that the photon
emitted/absorbed by a transition between the states has a
correspondingly low energy, or equivalently, long wavelength of about 21
cm. The decay from the excited (spins aligned) to the ground (spins
anti-aligned) state is very slow, with a mean lifetime of about 11
million years, but the ubiquity of hydrogen in the Universe makes the 21
cm, or ``HI'', line readily observable in emission (and absorption, but
this is not the focus of MARTINI) using radio telescopes. The fact that
the 21 cm radiation originates from a spectral line means that the
precise frequency in the rest frame is known from laboratory
measurements, so the observed frequency can be used to measure a Doppler
shift and therefore a measure of the kinematics of the emitting gas.
These properties lead to a wealth of scientific applications of HI line
emission observations, including in extragalactic astrophysics where
MARTINI focuses. A more detailed, forward-looking overview of the types
of applications in extragalactic HI radio astronomy that MARTINI is
suited to can be found in de Blok et al.
(\citeproc{ref-deBlok2015}{2015}).

Another powerful tool for the study of galaxies, this time coming from
the theory perspective, are simulations of galaxy formation. These come
in several flavours, one of which is ``hydrodynamical'' simulations
where gas is discretized as either a set of particles or a set of cells
in a mesh. Where particles are used, the most common formulation of the
relevant equations is ``smoothed particle hydrodynamics''
(\citeproc{ref-Gingold1977}{Gingold \& Monaghan, 1977}; SPH,
\citeproc{ref-Lucy1977}{Lucy, 1977}). Observations of the cosmic
microwave background can be used to constrain a multiscale Gaussian
random field to formulate initial conditions for a simulation of a
representative region of the early Universe, before the first galaxies
formed (e.g. \citeproc{ref-Bertschinger2001}{Bertschinger, 2001}). By
iteratively solving Poisson's equation for gravity, the hydrodynamic
equations, and additional equations and approximations needed to capture
relevant astrophysical processes, the simulated universe can be evolved
forward in time and make predictions for the structure and kinematics of
galaxies, including their atomic hydrogen gas.

\section{Statement of need}\label{statement-of-need}

A SPH simulation of galaxies and radio astronomical observations of
galaxies produce qualitatively different data sets. The output of a
simulation consists of tabulated lists of particles positions,
velocities, masses, chemical abundances, etc. A common high-level data
product from a radio observatory is a cubic data structure where the
axes of the cube correspond to right ascension (longitude on the sky),
declination (latitude) and frequency. Each cell in such a cube records
the intensity of radio emission with corresponding right ascension,
declination and frequency. Since two of the cube axes are spatial
coordinates and the third is a spectral coordinate, one can equivalently
think of these data cubes as images where each pixel contains a
discretized spectrum at that position (instead of a single-valued
intensity). Testing the theoretical predictions of simulations against
these kinds of observations is challenging because the data are
organized into fundamentally different structures. MARTINI provides a
tool to mimick the process of observing a simulated galaxy with a radio
telescope, producing as output the same kinds of ``data cubes'' and thus
enabling much more robust tests of the theoretical predictions made by
galaxy formation simulations.

\begin{figure}
\centering
\includegraphics{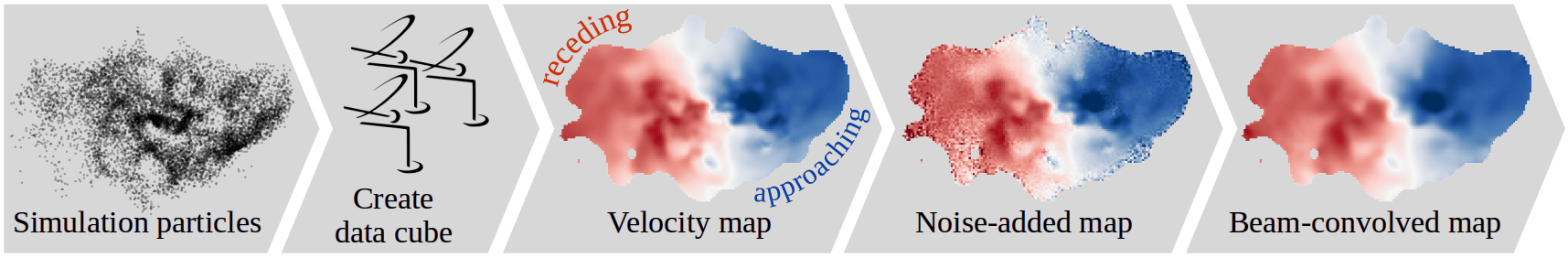}
\caption{MARTINI transforms a set of particles from a simulation into a
data cube as would be observed by a radio telescope. A velocity map is
one possible visualisation of a data cube and shows the mean Doppler
shift of the 21-cm radio emission line of atomic hydrogen in each pixel
of the cube as it approaches/recedes along its orbit in a plane inclined
to the observer. Properties of real radio observations such as noise and
the antenna response (the ``beam'') are faithfully mimicked.}
\end{figure}

The first cosmological hydrodynamical simulations that produced broadly
realistic populations of galaxies appeared around 2014
(\citeproc{ref-Schaye2015}{Schaye et al., 2015};
\citeproc{ref-Vogelsberger2014}{Vogelsberger et al., 2014}). A small
number of studies began to use mock observations of HI emission from
these simulated galaxies shortly after this
(\citeproc{ref-Read2016}{Read et al., 2016}), including the first work
using the code that would later become MARTINI
(\citeproc{ref-Oman2019}{Oman et al., 2019}). MARTINI has since been
adopted by the wider research community working on these types of
studies and is now used in the majority of such work including Chauhan
et al. (\citeproc{ref-Chauhan2019}{2019}), Mancera Piña et al.
(\citeproc{ref-ManceraPina2019}{2019}), Santos-Santos et al.
(\citeproc{ref-SantosSantos2020}{2020}), Mancera Piña et al.
(\citeproc{ref-ManceraPina2020}{2020}), Glowacki et al.
(\citeproc{ref-Glowacki2021}{2021}), Glowacki et al.
(\citeproc{ref-Glowacki2022}{2022}), Bilimogga et al.
(\citeproc{ref-Bilimogga2022}{2022}), Roper et al.
(\citeproc{ref-Roper2023}{2023}), Oman et al.
(\citeproc{ref-Oman2024}{2024}), and many more projects in progress.
Other codes that build on MARTINI have also started to appear, with
\href{https://github.com/MicheleDelliVeneri/ALMASim}{ALMASim}
(\citeproc{ref-Guglielmetti2023}{Guglielmetti et al., 2023}) as a first
example. While some authors have developed codes that implement some of
the same features as MARTINI (\citeproc{ref-Read2016}{Read et al.,
2016}), I am not aware of any such codes that have been publicly
released.

MARTINI is hosted on \href{https://github.com/kyleaoman/martini}{GitHub}
and has documentation available through
\href{https://martini.readthedocs.io}{ReadTheDocs}.

\section{Acknowledgements}\label{acknowledgements}

KAO acknowledges support by the Royal Society trhough Dorothy Hodgkin
Fellowship DHF/R1/231105, by STFC through grant ST/T000244/1, by the
European Research Council (ERC) through an Advanced Investigator Grant
to C. S. Frenk, DMIDAS (GA 786910), and by the Netherlands Foundation
for Scientific Research (NWO) through VICI grant 016.130.338 to M.
Verheijen. This work has made use of NASA's Astrophysics Data System.

\section*{References}\label{references}
\addcontentsline{toc}{section}{References}

\phantomsection\label{refs}


\begin{CSLReferences}{1}{0}
\bibitem[\citeproctext]{ref-Bertschinger2001}
Bertschinger, E. (2001). {Multiscale Gaussian Random Fields and Their
Application to Cosmological Simulations}. \emph{The Astrophysical
Journal Supplements}, \emph{137}(1), 1--20.
\url{https://doi.org/10.1086/322526}

\bibitem[\citeproctext]{ref-Bilimogga2022}
Bilimogga, P. V., Oman, K. A., Verheijen, M. A. W., \& van der Hulst, T.
(2022). {Using EAGLE simulations to study the effect of observational
constraints on the determination of H I asymmetries in galaxies}.
\emph{Monthly Notices of the Royal Astronomical Society}, \emph{513}(4),
5310--5327. \url{https://doi.org/10.1093/mnras/stac1213}

\bibitem[\citeproctext]{ref-Chauhan2019}
Chauhan, G., Lagos, C. del P., Obreschkow, D., Power, C., Oman, K., \&
Elahi, P. J. (2019). {The H I velocity function: a test of cosmology or
baryon physics?} \emph{Monthly Notices of the Royal Astronomical
Society}, \emph{488}(4), 5898--5915.
\url{https://doi.org/10.1093/mnras/stz2069}

\bibitem[\citeproctext]{ref-deBlok2015}
de Blok, E., Fraternali, F., Heald, G., Adams, B., Bosma, A., \&
Koribalski, B. (2015). {The SKA view of the Neutral Interstellar Medium
in Galaxies}. \emph{Advancing Astrophysics with the Square Kilometre
Array (AASKA14)}, 129. \url{https://doi.org/10.22323/1.215.0129}

\bibitem[\citeproctext]{ref-Gingold1977}
Gingold, R. A., \& Monaghan, J. J. (1977). {Smoothed particle
hydrodynamics: theory and application to non-spherical stars.}
\emph{Monthly Notices of the Royal Astronomical Society}, \emph{181},
375--389. \url{https://doi.org/10.1093/mnras/181.3.375}

\bibitem[\citeproctext]{ref-Glowacki2022}
Glowacki, M., Deg, N., Blyth, S.-L., Hank, N., Davé, R., Elson, E., \&
Spekkens, K. (2022). {ASymba: H I global profile asymmetries in the
SIMBA simulation}. \emph{Monthly Notices of the Royal Astronomical
Society}, \emph{517}(1), 1282--1298.
\url{https://doi.org/10.1093/mnras/stac2684}

\bibitem[\citeproctext]{ref-Glowacki2021}
Glowacki, M., Elson, E., \& Davé, R. (2021). {The redshift evolution of
the baryonic Tully-Fisher relation in SIMBA}. \emph{Monthly Notices of
the Royal Astronomical Society}, \emph{507}(3), 3267--3284.
\url{https://doi.org/10.1093/mnras/stab2279}

\bibitem[\citeproctext]{ref-Guglielmetti2023}
Guglielmetti, F., Delli Veneri, M., Baronchelli, I., Blanco, C., Dosi,
A., Enßlin, T., Johnson, V., Longo, G., Roth, J., Stoehr, F., Tychoniec,
Ł., \& Villard, E. (2023). {A BRAIN study to tackle image analysis with
artificial intelligence in the ALMA 2030 era}. \emph{arXiv e-Prints},
arXiv:2311.10657. \url{https://doi.org/10.48550/arXiv.2311.10657}

\bibitem[\citeproctext]{ref-Lucy1977}
Lucy, L. B. (1977). {A numerical approach to the testing of the fission
hypothesis.} \emph{The Astronomical Journal}, \emph{82}, 1013--1024.
\url{https://doi.org/10.1086/112164}

\bibitem[\citeproctext]{ref-ManceraPina2019}
Mancera Piña, P. E., Fraternali, F., Adams, E. A. K., Marasco, A.,
Oosterloo, T., Oman, K. A., Leisman, L., di Teodoro, E. M., Posti, L.,
Battipaglia, M., Cannon, J. M., Gault, L., Haynes, M. P., Janowiecki,
S., McAllan, E., Pagel, H. J., Reiter, K., Rhode, K. L., Salzer, J. J.,
\& Smith, N. J. (2019). {Off the Baryonic Tully-Fisher Relation: A
Population of Baryon-dominated Ultra-diffuse Galaxies}. \emph{The
Astrophysical Journal Letters}, \emph{883}(2), L33.
\url{https://doi.org/10.3847/2041-8213/ab40c7}

\bibitem[\citeproctext]{ref-ManceraPina2020}
Mancera Piña, P. E., Fraternali, F., Oman, K. A., Adams, E. A. K.,
Bacchini, C., Marasco, A., Oosterloo, T., Pezzulli, G., Posti, L.,
Leisman, L., Cannon, J. M., di Teodoro, E. M., Gault, L., Haynes, M. P.,
Reiter, K., Rhode, K. L., Salzer, J. J., \& Smith, N. J. (2020). {Robust
H I kinematics of gas-rich ultra-diffuse galaxies: hints of a
weak-feedback formation scenario}. \emph{Monthly Notices of the Royal
Astronomical Society}, \emph{495}(4), 3636--3655.
\url{https://doi.org/10.1093/mnras/staa1256}

\bibitem[\citeproctext]{ref-Oman2024}
Oman, K. A., Frenk, C. S., Crain, R. A., Lovell, M. R., \& Pfeffer, J.
(2024). {A warm dark matter cosmogony may yield more low-mass galaxy
detections in 21-cm surveys than a cold dark matter one}. \emph{arXiv
e-Prints}, arXiv:2401.11878.
\url{https://doi.org/10.48550/arXiv.2401.11878}

\bibitem[\citeproctext]{ref-Oman2019}
Oman, K. A., Marasco, A., Navarro, J. F., Frenk, C. S., Schaye, J., \&
Ben\'{i}tez-Llambay, A. (2019). {Non-circular motions and the diversity of
dwarf galaxy rotation curves}. \emph{Monthly Notices of the Royal
Astronomical Society}, \emph{482}(1), 821--847.
\url{https://doi.org/10.1093/mnras/sty2687}

\bibitem[\citeproctext]{ref-Read2016}
Read, J. I., Iorio, G., Agertz, O., \& Fraternali, F. (2016).
{Understanding the shape and diversity of dwarf galaxy rotation curves
in {\(\Lambda\)}CDM}. \emph{Monthly Notices of the Royal Astronomical
Society}, \emph{462}(4), 3628--3645.
\url{https://doi.org/10.1093/mnras/stw1876}

\bibitem[\citeproctext]{ref-Roper2023}
Roper, F. A., Oman, K. A., Frenk, C. S., Ben\'{i}tez-Llambay, A., Navarro,
J. F., \& Santos-Santos, I. M. E. (2023). {The diversity of rotation
curves of simulated galaxies with cusps and cores}. \emph{Monthly
Notices of the Royal Astronomical Society}, \emph{521}(1), 1316--1336.
\url{https://doi.org/10.1093/mnras/stad549}

\bibitem[\citeproctext]{ref-SantosSantos2020}
Santos-Santos, I. M. E., Navarro, J. F., Robertson, A., Ben\'{i}tez-Llambay,
A., Oman, K. A., Lovell, M. R., Frenk, C. S., Ludlow, A. D., Fattahi,
A., \& Ritz, A. (2020). {Baryonic clues to the puzzling diversity of
dwarf galaxy rotation curves}. \emph{Monthly Notices of the Royal
Astronomical Society}, \emph{495}(1), 58--77.
\url{https://doi.org/10.1093/mnras/staa1072}

\bibitem[\citeproctext]{ref-Schaye2015}
Schaye, J., Crain, R. A., Bower, R. G., Furlong, M., Schaller, M.,
Theuns, T., Dalla Vecchia, C., Frenk, C. S., McCarthy, I. G., Helly, J.
C., Jenkins, A., Rosas-Guevara, Y. M., White, S. D. M., Baes, M., Booth,
C. M., Camps, P., Navarro, J. F., Qu, Y., Rahmati, A., \ldots{}
Trayford, J. (2015). {The EAGLE project: simulating the evolution and
assembly of galaxies and their environments}. \emph{Monthly Notices of
the Royal Astronomical Society}, \emph{446}(1), 521--554.
\url{https://doi.org/10.1093/mnras/stu2058}

\bibitem[\citeproctext]{ref-Vogelsberger2014}
Vogelsberger, M., Genel, S., Springel, V., Torrey, P., Sijacki, D., Xu,
D., Snyder, G., Nelson, D., \& Hernquist, L. (2014). {Introducing the
Illustris Project: simulating the coevolution of dark and visible matter
in the Universe}. \emph{Monthly Notices of the Royal Astronomical
Society}, \emph{444}(2), 1518--1547.
\url{https://doi.org/10.1093/mnras/stu1536}

\end{CSLReferences}
\end{document}